\documentclass[notitlepage,twocolumn,prl,nobalancelastpage,amssymb,superscriptaddress,showpacs,preprintnumbers,nofootinbib]{revtex4-2}
\usepackage[utf8]{inputenc}
\usepackage[normalem]{ulem}
\usepackage[dvipsnames]{xcolor}
\usepackage{cancel}

\definecolor{red}{rgb}{0.9, 0,0}
\definecolor{cerulean}{rgb}{0., 0.42,0.9}
\definecolor{navy}{rgb}{0.05, 0.05,0.8}

\usepackage[colorlinks]{hyperref}
\hypersetup{
	colorlinks = true,
	linkcolor = red,
	urlcolor  = blue,
	citecolor = blue,
	anchorcolor = blue
}

\usepackage{amsmath,amsfonts,amsthm,amssymb, bm, mathtools, latexsym} 
\usepackage{hyperref}
\usepackage{subcaption}
\usepackage{physics}
\usepackage[dvipsnames]{xcolor}
\usepackage{enumerate}
\usepackage{slashed}
\usepackage[final]{showlabels}

\def\e{{\epsilon}}

\def\ka{{\kappa}}

\def\a{{\alpha}}
\def\b{{\beta}}
\def\g{{\gamma}}
\def\l{{\lambda}}
\def\G{{\Gamma}}
\def\Th{{\Theta}}
\def\d{{\delta}}

\def\O{\Omega}

\def\CB{{\mathcal B}}

\def\CH{{\mathcal H}}

\def\CO{{\mathcal O}}

\newcommand{\tu}{\tilde{u}}
\newcommand{\dt}{{\text d}}
\newcommand{\p}{\partial}

\newcommand{\Apart}{{\mu}}

\NewDocumentCommand{\codeword}{v}{%
	\texttt{\textcolor{blue}{#1}}%
}

\usepackage{subfiles} 

\begin{document}
	
	
	\title{Quantum Mechanics of a Spherically Symmetric Causal Diamond in Minkowski Spacetime}
	\author{Mathew W. Bub}
	\affiliation{Walter Burke Institute for Theoretical Physics, California Institute of Technology, Pasadena, CA 91125}
	\author{Temple He}
	\affiliation{Walter Burke Institute for Theoretical Physics, California Institute of Technology, Pasadena, CA 91125}
	\author{Prahar Mitra}
	\affiliation{Institute for Theoretical Physics, University of Amsterdam, Science Park 904, Postbus 94485, 1090 GL Amsterdam, The Netherlands}
	\author{Yiwen Zhang}
	\affiliation{Walter Burke Institute for Theoretical Physics, California Institute of Technology, Pasadena, CA 91125}
	\author{Kathryn M. Zurek}
	\affiliation{Walter Burke Institute for Theoretical Physics, California Institute of Technology, Pasadena, CA 91125}
	
	\begin{abstract}
		\noindent
		
		We construct the phase space of a spherically symmetric causal diamond in $(d+2)$-dimensional Minkowski spacetime. Utilizing the covariant phase space formalism, we identify the relevant degrees of freedom that localize to the $d$-dimensional bifurcate horizon and, upon canonical quantization, determine their commutators.  On this phase space, we find two Iyer-Wald charges. The first of these charges, proportional to the area of the causal diamond, is responsible for shifting the null time along the horizon and has been well-documented in the literature. The second charge is much less understood, being integrable for $d \geq 2$ only if we allow for field-dependent diffeomorphisms and is responsible for changing the size of the causal diamond.
	\end{abstract}
	
	\maketitle
	
	\preprint{CALT-TH 2024-032}
	
	\maketitle
	
	%
	%
	
	\noindent\textbf{Introduction.}
	Enormous conceptual progress has been made to unite gravity and quantum mechanics, and horizons have played a crucial role in these developments. In the presence of a horizon, there is a fairly universal notion of entropy, with the covariant entropy bound stating that the maximal entropy of any quantum state within a horizon is bounded above by $A/4 G_N$, where $A$ is the area of the horizon. This bound takes the same form as the Bekenstein-Hawking entropy. These results can be shown for conformal field theories with a gravity dual \cite{Ryu:2006ef, Casini:2011kv}, and it has been argued that they apply much more broadly to causal diamonds in maximally symmetric spacetimes \cite{Jacobson:2018ahi, Verlinde:2019ade, Banks:2021jwj}.  However, outside of the context of AdS/CFT, comparatively few computational tools are available to analyze the quantum mechanics of causal horizons created by lightsheets.  
	
	Despite this difficulty, it is essential to understand the role of quantum mechanics in the study of causal diamonds in Minkowski spacetime, which serves as an excellent approximation of the spacetime accessible by laboratory experiments. Fortunately, there has been much research into the algebra of observables of black hole horizons and, more generally, Killing horizons \cite{Carlip:1994gy, Carlip:1998wz, Carlip:1998qw, Solodukhin:1998tc, Carlip:1999cy, Silva:2002jq, Carlip:2011vr, Carlip:2017xne, Carlip:2019dbu}, and it has been argued that many of the tools developed are also applicable to the study of causal horizons \cite{Banks:2021jwj}. More recently, research along this direction has further been developed from the perspective of asymptotic symmetries \cite{Koga:2001vq, Hawking:2015qqa, Donnay:2015abr, Eling:2016xlx, Donnay:2016ejv, Hawking:2016sgy, Eling:2016qvx, Haco:2018ske, Grumiller:2019fmp, Adami:2020amw, Donnay:2020yxw, Adami:2021nnf, He:2024vlp}, the covariant phase space formalism \cite{Chandrasekaran:2018aop, Chandrasekaran:2019ewn, Chandrasekaran:2020wwn, Ciambelli:2023mir, Ciambelli:2024swv, Pulakkat:2024qps}, and von Neumann algebras \cite{Donnelly:2022kfs, Faulkner:2024gst, Ciambelli:2024qgi}. In many of these works, a Hamiltonian charge corresponding to the area of the horizon has been obtained~\cite{Carlip:1998wz, Solodukhin:1998tc, Carlip:1999cy, Donnay:2015abr, Donnay:2016ejv, Eling:2016qvx, Donnay:2020yxw, Chandrasekaran:2018aop, Chandrasekaran:2019ewn, He:2024vlp}.
	
	In this letter, we consider a relatively simple setup involving a spherically symmetric causal diamond in $(d+2)$-dimensional Minkowski spacetime.\footnote{Related analyses similar in spirit, but involving instead Jackiw–Teitelboim gravity, were done in~\cite{Harlow:2018tqv, Gukov:2022oed}.} The simplicity of this spacetime allows us to straightforwardly construct the symplectic form, which we invert to obtain the quantum commutators. Furthermore, we derive the associated Iyer-Wald Hamiltonian charges \cite{Iyer:1994ys}. One such charge is the area operator, which generates shifts in null time along the horizon. This is also the boost generator, which has been associated with the vacuum modular Hamiltonian of the system, and has been previously studied in similar contexts \cite{Jacobson:2018ahi, Chandrasekaran:2019ewn}. However, we discover that by allowing for field-dependent diffeomorphisms, a second (integrable) charge exists, which generates shifts in the global time coordinate and causes the causal diamond to shrink or expand. This pair of charges fully characterizes the quantum mechanics of a spherically symmetric causal diamond, and it would be extremely interesting to explore the observational implications of both charges, as well as their fluctuations, which we leave for future work.
	
	In this letter, we first construct a coordinate system that describes a causal diamond in Minkowski spacetime. We then derive the symplectic form and quantum commutators associated to the causal diamond via the covariant phase space formalism. Finally, we compute the two families of Iyer-Wald charges and elaborate on their physical significance.

	%
	%
	
	\medskip
	
	\noindent\textbf{Parametrizing the causal diamond.} In retarded coordinates, the metric of $(d+2)$-dimensional Minkowski spacetime is given by\footnote{Alternatively, we could have worked with advanced time instead. However, as we will see, the degrees of freedom completely localizes to the bifurcate horizon $\CB$, and hence both coordinate choices are equivalent.}
	\begin{align}\label{bondi-metric}
		\dt s^2 = -\dt \tu^2 - 2 \,\dt \tu\,\dt \tilde{r} + \tilde{r}^2 \dt \O_d^2 ,
	\end{align}
	where $\tu = {\tilde t} - {\tilde r}$ is the retarded time coordinate and $\dt \O_d^2$ is the round metric of the unit $S^d$. We consider a spherically symmetric causal diamond of size $L$, defined by $|{\tilde t}+L|+{\tilde r} \leq L$. In retarded coordinates this is given by $-2L \leq \tu \leq - 2 {\tilde r}$. The past and future null boundaries $\CH^\pm$ of the diamond are respectively given by $\tu = -2L$ and $\tu =  - 2 {\tilde r}$. The bifurcate horizon $\CB$, which we define to be the intersection between $\CH^+$ and $\CH^-$, is given by ${\tilde r}=L$ and $\tu=-2L$. We are interested in fluctuations of the causal diamond that arise from spherically symmetric (large) diffeomorphisms. To this end, we consider a coordinate transformation of the form
	\begin{align}\label{diffeo}
		\begin{split}
			\tu = - 2 \Phi_0(u) , \qquad \tilde{r} = \Phi(u,r) . 
		\end{split}
	\end{align}
	A diffeomorphism of this form ensures that level sets of $u$ are the same as those of $\tu$ and, therefore, null. We require $u$ to increase to the future and $r$ to increase inwards, so we take $\p_u \Phi_0<0$ and $\p_r \Phi < 0$. The new coordinates cover the entire interior of the diamond as long as $\Phi \leq \Phi_0 \leq L$. The past horizon $\CH^-$ is given by $u=u_-$ where
	\begin{equation}
		\begin{split}
			\label{um_def}
			\Phi_0(u_-)=L  .
		\end{split}
	\end{equation}
	We require that the future null boundary $\CH^+$ of the diamond be given by $r=0$, which implies 
	\begin{align}\label{Phi0-def}
		\Phi_0(u) = \lim_{r \to 0} \Phi(u,r).
	\end{align}
	The bifurcate horizon $\CB$ is then located at $r=0$, $u=u_-$. Note that $\CH^+$ is a Cauchy slice for the causal diamond, and the following symplectic analysis will be carried out on this surface.

\begin{figure} [h]
		\centering
		\includegraphics[scale=0.15]{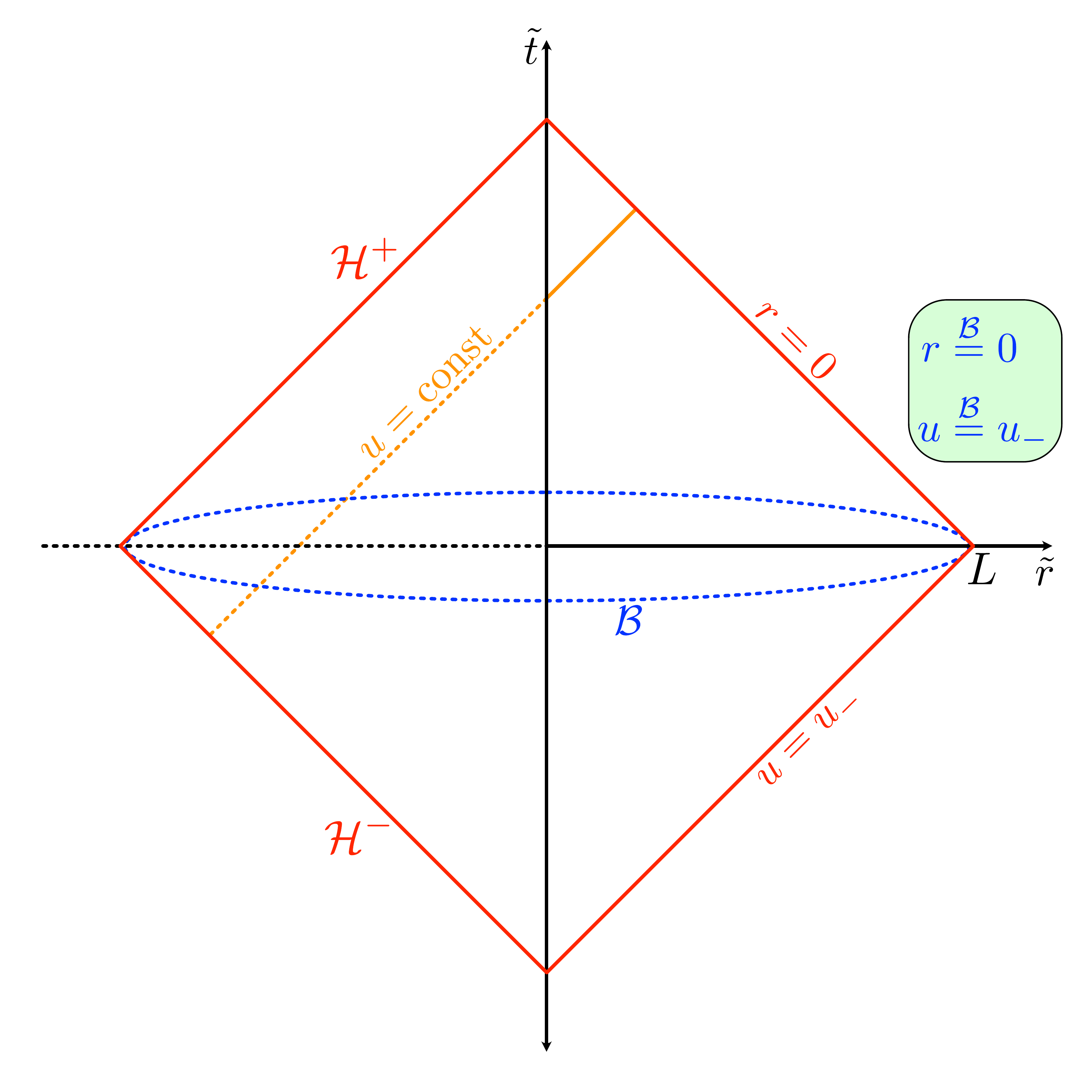}
		\caption{A causal diamond of radius $L$ with null boundaries shown as solid red line. The blue dashed ellipse represents the $d$-dimensional bifurcation surface (that can be integrated over in spherically symmetric solutions), while the constant $u$ hypersurface is represented by the orange line.}
		\label{fig:causal_diamond}
	\end{figure}

	In the new coordinates, the metric Eq.~\eqref{bondi-metric} takes the form
	\begin{align}\label{met-final}
		\begin{split}
			\dt s^2 &= -2 \ka(u,r)r e^{2\b(u,r)} \dt u^2 + 2 e^{2\b(u,r)} \dt u\, \dt r \\
			&\qquad + \Phi(u,r)^2 \dt \O_d^2,
		\end{split}
	\end{align}
	where 
	\begin{align}\label{constraints}
		\begin{split}
			e^{2\b(u,r)} &= 2 \p_u \Phi_0(u) \p_r\Phi(u,r), \\ \ka(u,r) &= \frac{\p_u \big( \Phi_0(u) - \Phi(u,r) \big)}{r\p_r\Phi(u,r)} .
		\end{split}
	\end{align}
	Although this metric resembles the Gaussian null coordinates used near null hypersurfaces, Eq.~\eqref{met-final} describes the entire causal diamond. For reference, a spacetime diagram of the causal diamond is shown in Fig.~\ref{fig:causal_diamond}.

	%
	%
	
	\medskip
	
	\noindent\textbf{Derivation of the symplectic form.} 
	We proceed to construct the symplectic form for the spherically symmetric causal diamond, via the use of the covariant phase space formalism \cite{Crnkovic:1986ex, Lee:1990nz, Iyer:1994ys, Wald:1999wa}. We vary the Einstein-Hilbert action, isolate the boundary term, and then integrate the latter over a Cauchy slice $\Sigma$ to obtain the pre-symplectic potential (e.g., see \cite{Carroll:2004st}). The result is\footnote{The tilde emphasizes that Eq.~\eqref{eq:potential_einstein} is the \emph{pre}-symplectic potential rather than the symplectic potential, as the corresponding \emph{pre}-symplectic form $\widetilde{\Omega}$ may not be invertible. The tilde is dropped once we gauge-fix in Eq.~\eqref{Phi_form} and Eq.~\eqref{ka_form}, which renders $\widetilde\Omega$ invertible.}
	\begin{equation} \label{eq:potential_einstein}
		\widetilde{\Theta}_\Sigma [ g ; \d g]  = \frac{1}{16 \pi G_N} \int_\Sigma \dt \Sigma_\mu \, (g^{\nu \rho} \d \Gamma^\mu_{\nu \rho} - g^{\mu \nu} \d \Gamma^\rho_{\nu \rho}) ,
	\end{equation}
	where $\dt \Sigma_\mu$ is the surface element on $\Sigma$. For the special case where $\Sigma = \mathcal{H}^+$, the surface element is given by\footnote{The negative sign in Eq.~\eqref{eq:measure} arises from the fact our normal vector is outward-pointing, and this is in the direction of decreasing $r$, since $r$ increases as we move \emph{into} the causal diamond along constant $u$ rays.}
	\begin{equation}\label{eq:measure}
		\dt \Sigma_\mu = -\delta_\mu^r \, \Phi^{d} e^{2\b} \, \dt u \, \dt \O_d ,
	\end{equation}
	where $\dt \O_d$ is the volume form on $S^d$. We derive in the Supplementary Material that given the definition
	\begin{align}\label{eq:varphi-def}
		\varphi(u,r) \equiv \Phi(u,r)^d, 
	\end{align}
	the pre-symplectic potential is given by 
	\begin{align}\label{eq:theta-fin}
		\begin{split}
			&\widetilde \Theta_{\CH^+} [ g ; \d g] =  \frac{\O_d}{8\pi G_N} ( \log|\p_u\varphi |  - \b ) \d \varphi \big|_\CB + \delta(\,\cdots) ,
		\end{split}
	\end{align}
	where $\O_d = \frac{2\pi^{\frac{d+1}{2}}}{\G(\frac{d+1}{2})}$ is the volume of $S^d$. We note that the pre-symplectic potential localizes entirely on the bifurcate horizon $\CB$ with no contribution from the rest of $\CH^+$. The reason is that in our rather simplistic setup, all metric fluctuations arise due to diffeomorphisms, and it was shown in \cite{Iyer:1994ys} that such contributions are a boundary term. The $\delta(\,\cdots)$ denotes the total variational terms that do not contribute to the pre-symplectic form, and are neglected henceforth.\footnote{The pre-symplectic potential in fact has further ambiguities arising from possible corner terms, as explained in Section 1.3.3 of \cite{Compere:2018aar}. However, due to spherical symmetry, all such local and covariant corner terms are total variations that do not affect the symplectic form.}

The pre-symplectic form is given by
	\begin{equation}
		\begin{split}
			\widetilde{\Omega} [ g; \delta_1 g , \delta_2 g] &= \d_1 \widetilde \Theta_\Sigma[g ; \d_2 g] - \d_2 \widetilde \Theta_\Sigma[g ; \d_1 g] \\
			&\qquad  - \widetilde \Theta_\Sigma [ g ; [ \d_1 , \d_2 ] g ] ,
		\end{split}
	\end{equation}
	where we dropped the subscript $\Sigma$ on $\widetilde\O$ since we will eventually see the symplectic form is independent of $\Sigma$. It then follows from Eq.~\eqref{eq:theta-fin} that
	\begin{align}\label{symp-form1}
		\begin{split}
			\widetilde{\O}[g ; \delta_1 g, \delta_2 g] &= \frac{\O_d}{8\pi G_N} \delta ( \log|\p_u\varphi |  - \b ) \wedge \delta \varphi \big|_\CB ,
		\end{split}
	\end{align}
	where we define the notation $\d a \wedge \d b \equiv \d_1 a \d_2 b - \d_2 a \d_1 b$. Furthermore, observing that 
	\begin{align}
		\begin{split}
			\big( \log|\p_u\varphi |  - \b \big) \big|_{\CB} = \bigg( \log \frac{\Phi_0^{d-1} d}{2} + \frac{1}{2} \log \frac{2 \p_u\Phi_0}{\p_r\Phi}   \bigg) \bigg|_{\CB} ,
		\end{split}
	\end{align}
	we can use the antisymmetry of the wedge product to obtain
	\begin{align}\label{eq:O-fin1}
		\begin{split}
			\widetilde{\O}[g ; \delta_1 g, \delta_2 g] &=  \frac{1}{8\pi G_N} \d \mu \wedge \d A ,
		\end{split}
	\end{align}
	where
	\begin{align}\label{eq:mu-def}
		A \equiv \O_d \varphi\big|_{\CB} = \O_d \Phi^d \big|_{\CB}, \qquad \Apart = \frac{1}{2} \log  \frac{2\p_u\Phi}{\p_r\Phi}  \bigg|_{\CB} .
	\end{align}
	
	To elevate this to the symplectic form, we demand invertibility and therefore need to perform a small gauge  fixing. This is done in the Supplementary Material, and we obtain
	\begin{equation}
		\begin{split}
			\b(u,r) = 0 , \qquad \ka(u,r) = \ka_0 ,
		\end{split}
	\end{equation}
	where $\ka_0$ is a spacetime constant, and 
	\begin{align}\label{Phi_form}
		\begin{split}
			\Phi(u,r) &= L - \frac{1}{2\ka_0} e^{\ka_0 u + \a} - r e^{-\ka_0 u - \a} ,
		\end{split}
	\end{align}
	where $\a$ is another spacetime constant. We also note that in this gauge, $u$ is the proper time for a uniformly accelerating observer at $r = (2 \ka_0)^{-1}$, with their proper acceleration $a=\ka_0$ and whose future Rindler horizon is $\CH^+$. The Unruh temperature experienced by this observer is therefore
	\begin{equation}
		\begin{split}\label{Unruh_Temp}
			T = \frac{\ka_0}{2\pi} . 
		\end{split}
	\end{equation}
	
	We now evaluate the symplectic form in this gauge. From the definitions Eq.~\eqref{eq:mu-def}, it follows that
	\begin{align}\label{eq:defs-final}
		A = \O_d L^d , \qquad \mu = \ka_0 u_- + \a.
	\end{align}
	From Eqs.~\eqref{um_def}, \eqref{Phi0-def}, and \eqref{Phi_form}, it follows that $u_- = -\infty$, which implies that $\mu$, and therefore $\widetilde{\O}$, is generically formally divergent. In the Supplementary Material, we show that $\widetilde\O$ can be made finite and invertible by assuming that $\ka_0$ is non-dynamical on the phase space and is instead fixed in terms of $L$
	\begin{equation}
		\begin{split}
			\label{ka_form}
			\ka_0 \equiv \ka_0(L) .
		\end{split}
	\end{equation}
	With this assumption, the symplectic form reduces to
	\begin{align}\label{eq:symp-form-final}
		\begin{split}
			\O[g ; \delta_1 g, \delta_2 g] = \frac{1}{8\pi G_N} \delta \a \wedge \delta A . 
		\end{split}
	\end{align}
Inverting Eq.~\eqref{eq:symp-form-final}, we immediately arrive at the Poisson bracket
	\begin{align}
		\begin{split}
			\{ \a , A \} = - 8\pi G_N .
		\end{split}
	\end{align}
	Upon canonical quantization, we promote the Poisson bracket to a quantum commutator so that
	\begin{align}\label{eq:quan-comm}
		[ \a , A ] = - 8\pi i G_N.
	\end{align}

	%
	%
	
	\medskip
	
	\noindent\textbf{Hamiltonian charges.} We now derive the Iyer-Wald Hamiltonian charges corresponding to horizon-preserving diffeomorphisms of the causal diamond. Recall that for a generic diffeomorphism generated by a vector field $\xi^\mu$, the Iyer-Wald Hamiltonian charge $H_\xi$ is obtained from the symplectic form via the variational equation \cite{Iyer:1994ys}
	\begin{equation}\label{IW-charge}
		\slashed{\delta} H_\xi = \Omega[ \phi ; \d \phi , \delta_\xi \phi ]  . 
	\end{equation}
	This is merely Hamilton's equation expressed in terms of the covariant symplectic form. The slash on the left-hand side indicates that the right-hand side might not be a total variation, and hence $H_\xi$ may not exist. If $H_\xi$ does exist, Eq.~\eqref{IW-charge} is integrable, and the charge is obtained by inverting the variation.
	
	We consider diffeomorphisms that preserve the form of our metric Eq.~\eqref{met-final}, with $\b$ gauge-fixed to $\b=0$ and $\ka(u,r) = \ka_0$. This amounts to requiring the following conditions on the vector field $\xi^\mu$ generating the diffeomorphism:
	\begin{equation} \label{eq:lie_derivatives}
		\mathcal{L}_\xi g_{uu} = \mathcal{O}(r), \quad \mathcal{L}_\xi g_{ur} = \mathcal{L}_\xi g_{rr} = 0.
	\end{equation}
	Symmetries of this or related forms have been studied numerous times in the literature, most famously by Bondi, van der Burg, Metzner, and Sachs \cite{Bondi:1962px, Sachs:1962wk, Sachs:1962zza} in the context of asymptotically flat spacetimes. More recently, asymptotic symmetries of null surfaces have been the subject of significant study \cite{Carlip:1998wz,Carlip:1999cy,Donnay:2015abr,Donnay:2016ejv,Chandrasekaran:2020wwn,Chandrasekaran:2018aop,Chandrasekaran:2019ewn,Grumiller:2019fmp,Adami:2020amw,Donnay:2020yxw, Adami:2021nnf, He:2024vlp}. In our case, we also require that $\xi^\mu$ preserves the spherical symmetry of the diamond, which significantly simplifies our calculation. The resultant set of allowed diffeomorphisms is generated by
	\begin{equation}
		\xi^\mu = ( f(u), -r \p_u f(u), \vec 0 )  ,
	\end{equation}
	where $f(u)$ is any smooth function of $u$. Under an infinitesimal diffeomorphism $x^\mu \to x^\mu + \xi^\mu$, we have
	\begin{align}
		\begin{split}
			\d_\xi \ka_0 = - \l_{\ka} ~ &\implies ~ \ka_0 \p_u f + \p_u^2 f =  \l_\ka \\
			&\implies~ f(u) = c_1 + c_2 e^{-\ka_0 u - \a} + \frac{\l_\ka u}{\ka_0} ,
		\end{split}	
	\end{align}
	where $c_1,c_2$ are some undetermined constants and $\l_\ka$ is an infinitesimal parameter. If we further fix
	\begin{align}
		c_1 = \l_\a - \frac{\l_\ka}{\ka_0^2} , \qquad c_2 = - 2 \l_L,
	\end{align}
	for infinitesimal parameters $\l_\a$ and $\l_L$, then the diffeomorphism generated by $\xi^\mu$ yields
	\begin{align}\label{eq:variations}
		\d_\xi \ka_0 = - \l_\ka , \qquad \d_\xi L = - \l_L, \qquad \d_\xi \a = - \ka_0 \l_\a.
	\end{align}
	Recall that we require $\ka_0 \equiv \ka_0(L)$ in order for the symplectic form Eq.~\eqref{eq:symp-form-final} to be finite, which implies
	\begin{align}
		\l_\ka = \ka'_0(L) \l_L.
	\end{align}
Comparing this with the field $\Phi$ given in Eq.~\eqref{Phi_form}, we immediately see that $\l_\a$ parametrizes the shift in null time $u$, whereas $\l_L$ parametrizes the change in the size of the causal diamond $L$. Thus, the most general symmetry preserving the geometry of a spherically symmetric causal diamond is parametrized by $\l_L$ and $\l_\a$. We now construct the Iyer-Wald charges that correspond to the two diffeomorphisms above.

	First, let us construct the charge where $\l_\a \not= 0$ but $\l_L = 0$. This is the charge that generates null time shifts in $u$ but keeps $L$ fixed. Using Eq.~\eqref{IW-charge} with the symplectic form Eq.~\eqref{eq:symp-form-final}, we obtain
	\begin{align}
		\begin{split}
			\slashed\d H_{\a} &= - \frac{1}{8\pi G_N} \d_\xi \a \d A \\
			&= \d \bigg( \frac{\l_\a \Omega_d d}{8\pi G_N} \int_0^L \dt L'\, \ka_0(L') L'^{d-1} \bigg),
		\end{split}
	\end{align}
	where we used Eq.~\eqref{eq:defs-final} and Eq.~\eqref{eq:variations}. We conclude that the first charge is
	\begin{align}
		\label{Ha_def}
		H_\a = \frac{\O_d d}{8\pi G_N} \int_0^L \dt L'\, \ka_0(L') L'^{d-1} ,
	\end{align}
where we normalized the charge $H_\a$ to exclude $\l_\a$. When there are no other length scales in the theory, dimensional analysis implies that $\ka_0$ takes the form $\ka_0(L) = C/ L$ for some dimensionless constant $C$ independent of $L$. In this case, the charge above is given by\footnote{In three dimensions ($d=1$), we have $H_\a = \frac{C}{4 G_N} \log L $.}
	\begin{equation}
		\begin{split}
			H_\a = \frac{d}{d-1} \frac{\O_d}{8\pi G_N} C L^{d-1} = \frac{d}{d-1} \frac{\ka_0(L)}{2\pi} \frac{A}{4G_N}.
		\end{split}
	\end{equation}
	Identifying $\ka_0(L) /2\pi$ with the Unruh temperature of an accelerating observer (see Eq.~\eqref{Unruh_Temp}) and associating an entropy $S = A/4G_N$ to the causal diamond, we find a version of the Smarr formula for causal diamonds:
	\begin{equation}
		\begin{split}
			\frac{d-1}{d} H_\a = T S . 
		\end{split}
	\end{equation}
	From Eq.~\eqref{Ha_def}, we can also reproduce a version of the first law for causal diamonds, namely\footnote{Notice that because the vector generating shifts in $u$ is timelike, the energy measured by the accelerating observer is $E = -H_\a$, which implies that the temperature associated to causal diamonds is negative. This is consistent with the results of \cite{Jacobson:2018ahi, Chandrasekaran:2019ewn}.}
	\begin{equation}
		\begin{split}
			\label{first_law_cd}
			\d H_\a &= \frac{\O_d d}{8\pi G_N} \ka_0(L) L^{d-1} \d L = \frac{\ka_0(L)}{2\pi} \d \left( \frac{A}{4G_N} \right) = T \d S .
		\end{split}
	\end{equation}
	
	Note that $H_\a$ generates time translations in $u$, which is the proper time of a uniformly accelerating observer. Therefore, we conclude that $H_\a$ is in fact the generator of boosts that preserve the causal diamond and is also the vacuum modular Hamiltonian of the causal diamond (see e.g., \cite{Bisognano:1975, Bisognano:1976}).

	Next, we construct the second charge where $\l_\a = 0$ but $\l_L \not= 0$. Using Eq.~\eqref{IW-charge} with the symplectic form Eq.~\eqref{eq:symp-form-final}, we obtain 
	\begin{align}
		\begin{split}
			\slashed\d H_L &= \frac{1}{8\pi G_N} \delta\a \d_\xi A = - \frac{\l_L \O_d d}{8\pi G_N} L^{d-1} \delta\a ,
		\end{split}
	\end{align}
	where we again used Eq.~\eqref{eq:defs-final} and Eq.~\eqref{eq:variations}. Because $\d L \not = 0$ in general, the above equation is not a total variation, which implies that the charge is not integrable.\footnote{The exception to this is if $d=1$ and we are in a three-dimensional spacetime, which was observed in \cite{Donnay:2016ejv}. In that case $L^{d-1}=1$ and there are no obstructions to writing down the charge.} However, for the special case where we consider a field-dependent diffeomorphism and fix
	\begin{align}\label{field-dep-shift}
		\l_L = \frac{G_N \e}{L^{d-1}} ,
	\end{align}
where $\e$ is an infinitesimal parameter, $\l_L L^{d-1}$ becomes a constant on phase space. Thus, the charge becomes integrable and is given by
	\begin{align}\label{eq:charge2}
		H_L = - \frac{\O_d d}{8\pi} \a,
	\end{align}
where we normalized the charge to exclude $\e$. As we previously mentioned, this charge generates a shift in the size of the causal diamond $L$, but the shifts are field-dependent, given by Eq.~\eqref{field-dep-shift}. However, the change in the area of the diamond is field-independent and is given by
	\begin{equation}
		\begin{split}
			\d_{\xi} A &= - \O_d L^{d-1} \l_L d = - \O_d G_N d ,
		\end{split}
	\end{equation}
	where we used \eqref{field-dep-shift}.

\begin{figure} [h]
		\centering
		\includegraphics[scale=0.15]{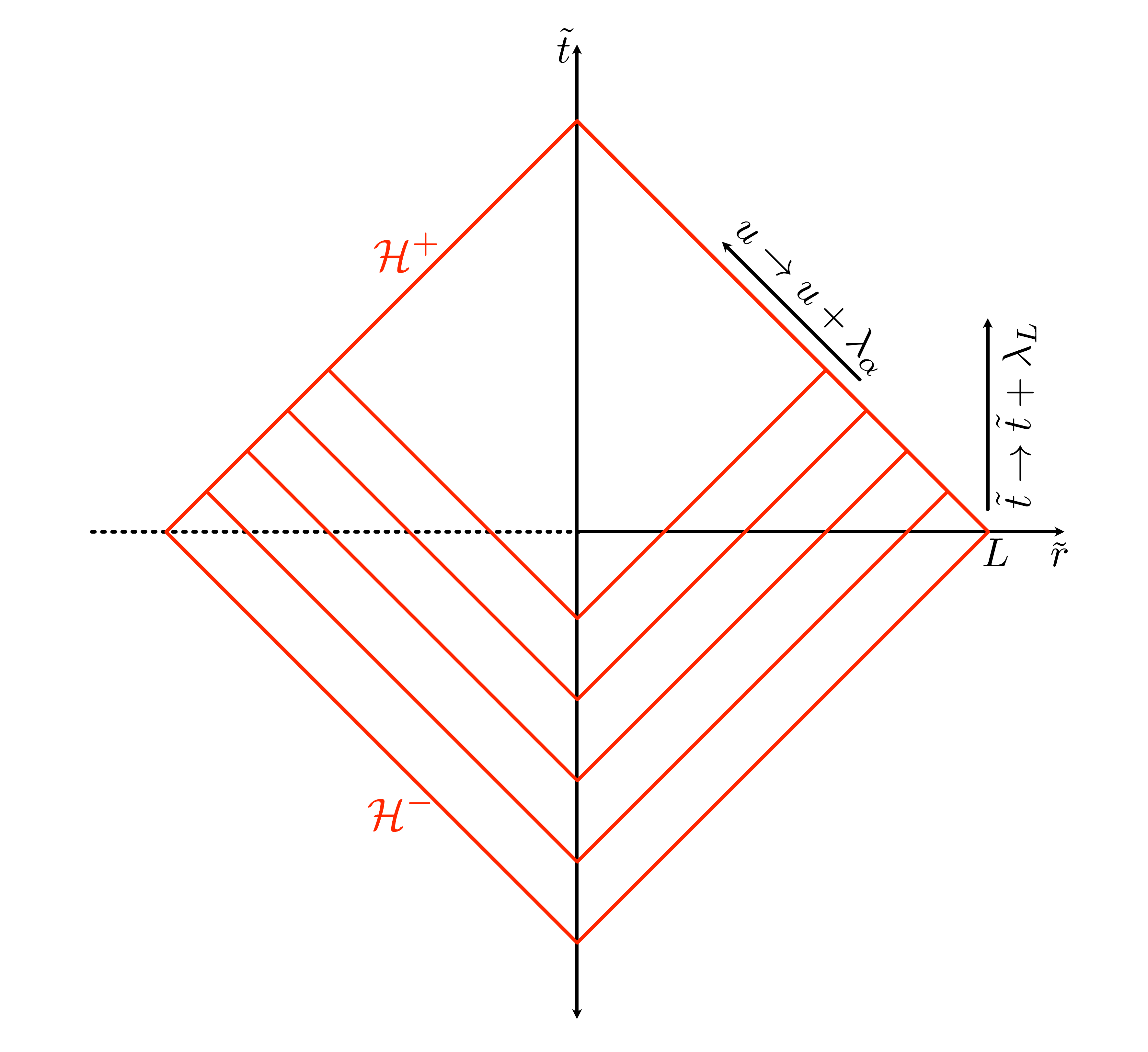}
		\caption{Pictorial depiction of a series of ``nested'' causal diamonds. Charge $H_{\alpha}$ generates translation in null time $u \mapsto u + \l_\a$, whereas charge $H_L$ generates shift in the size of the diamond, which corresponds to, at the bifurcate horizon $\CB$, a translation in Minkowski time $\tilde{t} \mapsto \tilde{t} + \l_L$. } 
		\label{fig:nested_CD}
	\end{figure}

	There are some interesting observations regarding $H_L$. We note that while $H_L$ generates shifts in the size of the causal diamond, it additionally generates shifts in the global Minkowski time $\tilde{t}$. To see this, note that under the diffeomorphism generated by $\xi^\mu$ with $\l_\a = 0$, we have
	\begin{align}
		\begin{split}
			\Phi \to \Phi + \d_\xi \Phi &= L - \l_L - \frac{1}{2(\ka_0 - \l_\ka)} e^{(\ka_0 - \l_\ka) u + \a} \\
			&\qquad - r e^{-(\ka_0 - \l_\ka) u - \a} .
		\end{split}
	\end{align}
	Recalling that Minkowski time is given by $\tilde t = \tilde u + \tilde r$, we have to linear order in the variation using Eq.~\eqref{diffeo}
	\begin{align}
		\begin{split}
			\tilde t &\to \tilde t +  \l_L + \l_\ka \bigg( \frac{1}{2\ka_0^2}e^{\ka_0 u + \a} (1 - \ka_0 u) -  ru e^{-\ka_0 u - \a} \bigg) ,
		\end{split}
	\end{align}
	implying that $\tilde t$ indeed shifts under the action by $H_L$. In fact, at the bifurcate horizon $\CB$, the second term in the above expression vanishes as $u=-\infty$ and $r=0$, and we have $\tilde t \to \tilde t + \l_L$. Furthermore, recalling that the future boundary $\CH^+$ is located at $\tilde u + 2 \tilde r = 0$, it is straightforward to check that $\CH^+$ remains invariant under the action by $H_L$. On the other hand, the past horizon $\CH^-$ is located at $\tu = -2 L$, and this shifts under $\CH^-$ to $\tu = - 2 L + 2\l_L$. In practice, this occurs by shifting the location of the bifurcation surface while keeping the top of the diamond fixed, thereby creating a ``nesting'' of causal diamonds (see Fig.~\ref{fig:nested_CD}) \cite{Banks:2021jwj, Zurek:2022xzl, Zhang:2023mkf,Bak:2024kzk}. Thus, while $H_\a$ generates null time translations, $H_L$ acts like a dilatation operator (while keeping $\CH^+$ invariant). We can easily compute the commutation relation of the two charges using Eq.~\eqref{eq:quan-comm} to obtain
	\begin{align}
		[H_\a , H_L ] &= - \frac{i}{8\pi} \ka_0(L) \O_d d .
	\end{align}

	%
	%
	
	\medskip
	
	\noindent\textbf{Conclusion.} In this letter, we considered the quantum mechanics of a spherically symmetric causal diamond in $(d+2)$-dimensional spacetime. We constructed the symplectic form associated with the causal diamond, which localizes to the codimension-2 bifurcate horizon at the corner and also derived two physically relevant Hamiltonian charges using the Iyer-Wald prescription. The first charge $H_{\a}$ shifts the null time $u$ and is proportional to the area of the causal diamond. This is the vacuum modular Hamiltonian and has been extensively studied in the literature in a variety of contexts \cite{Bisognano:1975,Bisognano:1976,Casini:2011kv, Banks:2021jwj, Verlinde:2022hhs, Jacobson:2015hqa, Jacobson:2018ahi}. However, there is an interesting second charge $H_L$, which changes the size of the causal diamond, and it exists only if field-dependent diffeomorphisms (for $d \geq 2$) are allowed. Because our setup is simple enough to directly relate the original Minkowski spacetime to our Gaussian null coordinates, we can deduce that $H_L$ equivalently shifts the global Minkowski time.
	
	There are many natural future directions to explore. Perhaps first and foremost is to relax the spherical symmetry assumption. Many previous works have considered the full asymptotic symmetry algebra of generic null surfaces \cite{Carlip:1998wz,Carlip:1999cy,Donnay:2015abr,Donnay:2016ejv,Chandrasekaran:2020wwn,Chandrasekaran:2018aop,Chandrasekaran:2019ewn,Grumiller:2019fmp,Donnay:2020yxw, Adami:2020amw,Adami:2021nnf, Pulakkat:2024qps, He:2024vlp}, but have obtained differing results due to the choice of boundary conditions, charge integrability, and a variety of other challenges. Moreover, many of these works did not allow for field-dependent diffeomorphisms (notable exceptions are \cite{Grumiller:2019fmp, Adami:2021nnf, Pulakkat:2024qps, He:2024vlp} and references therein), which we have shown are essential for the existence of the second charge that generates shifts in the area of the bifurcation surface. To our knowledge, no analysis has been carried out to study this charge or its higher modes in a more general environment. It would be extremely interesting to understand whether the algebra of these higher modes admits a central extension, as well as to explore the observational consequences of the two charges. 
	
	%
	%
	
	\noindent\textbf{Acknowledgments.}
	We would like to thank Luca Ciambelli, Laura Donnay, Laurent Freidel, Antony Speranza, and C\'{e}line Zwikel for useful discussions. M.B., T.H., Y.Z., and K.Z. are supported by the Heising-Simons Foundation “Observational Signatures of Quantum Gravity” collaboration grant 2021-2817, the U.S. Department of Energy, Office of Science, Office of High Energy Physics, under Award No. DE-SC0011632, and the Walter Burke Institute for Theoretical Physics. M.B., Y.Z., and K.Z. are also supported by GQuEST funding provided from the U.S. Department of Energy via FNAL: DE-AC02-07CH11359. M.B. acknowledges the support of the Natural Sciences and Engineering Research Council of Canada (NSERC), [funding reference number PGS D - 578032 - 2023]. P.M. is supported by the European Research Council (ERC) under the European Union’s Horizon 2020 research and innovation programme (grant agreement No 852386). The work of K.Z. is also supported by a Simons Investigator award.

	\bibliography{references_use.bib}
	
	\clearpage
	\onecolumngrid
	\begin{center}
		\textbf{\large SUPPLEMENTARY MATERIAL \\[.2cm] ``The Quantum Mechanics of a Spherically Symmetric Causal Diamond in Minkowski Spacetime''}\\[.2cm]
		\vspace{0.05in}
		{Mathew W. Bub, Temple He, Prahar Mitra, Yiwen Zhang, and Kathryn M. Zurek}
	\end{center}
	
	\setcounter{equation}{0}
	\setcounter{figure}{0}
	\setcounter{table}{0}
	\setcounter{section}{0}
	\setcounter{page}{1}
	\makeatletter
	\renewcommand{\theequation}{S\arabic{equation}}
	\renewcommand{\thefigure}{S\arabic{figure}}
	\renewcommand{\thetable}{S\arabic{table}}
	
	\onecolumngrid
	
	\section{Deriving the Pre-symplectic Potential}
	\label{app:presymp-pot}
	
	In this section, our goal is to derive the pre-symplectic potential for the metric Eq.~\eqref{met-final} and obtain Eq.~\eqref{eq:theta-fin}. We begin by recalling Eq.~\eqref{eq:potential_einstein} and Eq.~\eqref{eq:measure}, which immediately implies
	\begin{align}\label{theta-def2}
		\begin{split}
			\widetilde \Th_{\CH^+}[g; \d g] &=  - \frac{1}{16\pi G_N} \int_{\CH^+} \dt u\, \dt \O_d \, e^{2\b} \Phi^d \Big[ g^{\mu\nu} (\delta\G^r_{\mu\nu}) - g^{\mu r}(\d \G^\l_{\l\mu} ) \Big] \\
			&= - \frac{1}{16\pi G_N} \int_{\CH^+} \dt u\,  \dt \O_d  \, \varphi \bigg[ 2 \delta\G^r_{ur}  + \frac{e^{2\b}}{\varphi^\frac{2}{d}} \g^{ab} \delta\G^r_{ab} - \d \G^\l_{\l u}  \bigg]  ,
		\end{split}
	\end{align}
	where we defined
	\begin{align}\label{varphi-def}
		\varphi(u,r) \equiv \Phi(u,r)^d .
	\end{align}
	Evaluating the variations of the Christoffel symbols, the integrand of Eq.~\eqref{theta-def2} becomes
	\begin{align}\label{theta-4}
		\begin{split}
			&\varphi \bigg[ 2 \delta\G^r_{ur}  + \frac{e^{2\b}}{\varphi^\frac{2}{d}} \g^{ab} \delta\G^r_{ab} - \d \G^\l_{\l u}  \bigg]  = - 2 \varphi \d \ka + 2 \p_u\varphi \delta\b  - 2 \bigg( \frac{1}{d}-1 \bigg) \frac{1}{\varphi}\p_u \varphi \delta\varphi - 2 \p_u\delta\varphi - 2\varphi \p_u \delta\b + \CO(r) .
		\end{split}
	\end{align}
	Now, recall that the pre-symplectic form is defined as
	\begin{align}\label{symp-form-def}
		\widetilde{\Omega} [ g; \delta_1 g , \delta_2 g] &= \d_1 \widetilde \Theta_\Sigma[g ; \d_2 g] - \d_2 \widetilde \Theta_\Sigma[g ; \d_1 g]   - \widetilde \Theta_\Sigma [ g ; [ \d_1 , \d_2 ] g ]  ,
	\end{align}
	so by the antisymmetry of $\d_1$ and $\d_2$, terms that are total variations in the pre-symplectic potential will not contribute to the pre-symplectic form, thereby allowing us to ignore such terms. Thus, we can rewrite Eq.~\eqref{theta-4} as
	\begin{align}
		\begin{split}
			&\varphi \bigg[ 2 \delta\G^r_{ur}  + \frac{e^{2\b}}{\varphi^\frac{2}{d}} \g^{ab} \delta\G^r_{ab} - \d \G^\l_{\l u}  \bigg]  =  \bigg[ 2 \ka + 4 \p_u \b - 2 \bigg( \frac{1}{d}-1 \bigg) \frac{1}{\varphi} \p_u\varphi \bigg] \delta\varphi - 2\p_u \big( \delta\varphi \b \big) + \delta(\cdots) + \CO(r) ,
		\end{split}
	\end{align}
	where $\delta(\cdots)$ indicate terms that are total variations. Substituting this back into Eq.~\eqref{theta-def2} and dropping the $\CO(r)$ terms since $r \to 0$ on $\CH^+$, we see that the pre-symplectic potential is given by
	\begin{align}
		\begin{split}
			\widetilde \Th_{\CH^+}[g; \d g] &= - \frac{\O_d}{16\pi G_N} \int_{u_-}^{u_+} \dt u  \bigg\{ \bigg[ 2 \ka + 4 \p_u \b - 2 \bigg( \frac{1}{d}-1 \bigg) \frac{1}{\varphi} \p_u\varphi \bigg] \delta\varphi - 2\p_u \big( \delta\varphi \b \big) \bigg\} + \delta(\cdots) ,
		\end{split}
	\end{align}
	where we integrated over the angular directions to obtain the volume of the unit $d$-sphere $\O_d$, and $u_\pm$ indicates the integration limits of the future horizon $\CH^+$. We can further simplify this by substituting it in Eq.~\eqref{constraints} to obtain
	\begin{align}\label{theta-5}
		\begin{split}
			\widetilde \Th_{\CH^+}[g; \d g] &=  - \frac{\O_d}{8\pi G_N} \int_{u_-}^{u_+} \dt u  \Big[ \p_u \log|\p_u \varphi| \delta\varphi - \p_u \big( \delta\varphi \b \big) \Big] + \delta(\cdots)  .
		\end{split}
	\end{align}
	We now note the first term in the integrand of Eq.~\eqref{theta-5} can be written as
	\begin{align}
		\begin{split}
			\p_u \log |\p_u \varphi|  \delta\varphi &= \p_u \big( \log|\p_u \varphi| \delta\varphi \big) - \log|\p_u\varphi| \p_u \delta\varphi  = \p_u \big( \log|\p_u \varphi| \delta\varphi \big) + \delta( \cdots),
		\end{split}
	\end{align}
	which means we can further simplify Eq.~\eqref{theta-5} to be
	\begin{align}\label{theta-fin}
		\begin{split}
			\widetilde \Th_{\CH^+}[g; \d g] &=   \frac{\O_d}{8\pi G_N} \big( \log|\p_u\varphi| - \b\big) \d \varphi \bigg|_{u = u_-} + \delta(\cdots)  ,
		\end{split}
	\end{align}
	where we carried out the $u$ integral, and noted that $\varphi(u_+) = \Phi(u_+,0)^d = 0$ since by definition $(u=u_+,r=0)$ is at the top of causal diamond, which corresponds to $\tilde{r} = 0$, and hence $\Phi(u_+,0) = 0$ by Eq.~\eqref{diffeo}. This is precisely Eq.~\eqref{eq:theta-fin}.

	\section{Promoting the Pre-symplectic Form to Symplectic Form}
	\label{app:invert-symp-form}
	
	In this section, we show how to gauge-fix so that the pre-symplectic form Eq.~\eqref{eq:O-fin1}, reproduced here for convenience as
	\begin{align}\label{eq:O-fin1-app}
		\widetilde\O[g; \d_1 g, \d_2 g] = \frac{1}{8\pi G_N} \d \mu \wedge \d A,
	\end{align}
	becomes invertible and hence the symplectic form Eq.~\eqref{eq:symp-form-final}. First, note that the pre-symplectic form written in the form Eq.~\eqref{symp-form1} does not depend on $\b$ independently, but rather depends on the linear combination $\log|\p_u \varphi| - \b$ at $\CB$. Hence, we can choose to gauge-fix $\b$, and since the symplectic form is independent of the value of $\b$ away from $\CB$, we choose the simplest choice, namely
	\begin{align}\label{app:beta-gauge }
		\b =  0 .
	\end{align}
	Recalling from Eq.~\eqref{constraints} the definition of $\b$, and we have upon setting $\b = 0$ in Eq.~\eqref{constraints} the equality
	\begin{align}\label{app:Phi1-def}
		\begin{split}
			\Phi(u,r) = \Phi_0(u) + \frac{r}{2\p_u \Phi_0(u)} . 
		\end{split}
	\end{align}
	Using Eq.~\eqref{constraints} and Eq.~\eqref{app:Phi1-def}, we similarly derive 
	\begin{align}\label{ka-def}
		\begin{split}
			\ka(u,r) &= \frac{\p_u^2\Phi_0(u)}{\p_u\Phi_0(u)} .
		\end{split}
	\end{align}
	
	Next, we note that the pre-symplectic form only depends on $\ka(u_-,0)$, which means it does not matter how $\ka$ changes as a function of $u$ and $r$. This allows to further gauge-fix so that $\ka(u, r) \equiv \kappa_0$ is independent of $u$ and $r$ and therefore a spacetime constant. In this case, we solve Eq.~\eqref{ka-def} to obtain \begin{align}\label{app:Phi_form}
		\begin{split}
			\Phi(u,r) = L - \frac{1}{2\ka_0} e^{\ka_0 u + \a} - r e^{-\ka_0 u - \a},
		\end{split}
	\end{align}
	where $L$ is the radius of the causal diamond, and $\a$ an arbitrary spacetime constant. Note that in our original Bondi coordinates Eq.~\eqref{bondi-metric}, the bifurcate horizon $\CB$ is located at $\tu = - 2 L$ and $\tilde r = L$, or equivalently at $u = u_-$ and $r = 0$. The coordinate transformation Eq.~\eqref{diffeo} suggests that $\CB$ is located at
	\begin{align}
		\tilde r = L = \Phi(u_-,0) = \Phi_0(u_-).
	\end{align}
	Comparing this to Eq.~\eqref{app:Phi_form}, we see that $\CB$, which also corresponds to the past boundary of $\CH^+$, is located at $u_- = -\infty$. We can also determine $u_+$ by using $\Phi(u_+,0) = 0$, which yields
	\begin{align}
		u_+ =  \frac{1}{\ka_0} \big( \log (2\ka_0 L) - \a \big).
	\end{align}
	
	Recalling Eq.~\eqref{eq:mu-def}, we see from Eq.~\eqref{app:Phi_form} that
	\begin{align}\label{app:defs-final}
		\begin{split}
			A = \O_d L^d, \qquad \mu = \lim_{u \to u_-} \big( \a + \ka_0 u \big) = \a + \ka_0 u_- . 
		\end{split}
	\end{align}
	Because $u_- = -\infty$, $\mu$ is formally divergent. Furthermore, from \eqref{app:defs-final}, it appears that there are three degrees of freedom, namely $L,\a,$ and $\ka_0$. Since the phase space is even-dimensional, it must be the case that not all three degrees of freedom are independent. We can resolve both issues if we assume $\ka_0 = \ka_0(L)$, so that $\ka_0$ depends purely on $L$ and is not an independent degree of freedom. In this case, $\ka_0$ is still a spacetime constant as it is independent of $u,r$, and we have $\d \ka_0 = \ka'(L) \d L$. By the antisymmetry of the wedge product, the divergent term vanishes, and we are left with\footnote{Another common choice in the literature has been to take $\kappa_0$ to be a fixed constant on the phase space, such that $\d \kappa_0 = 0$ (e.g., see \cite{Donnay:2015abr,Donnay:2016ejv,Eling:2016qvx,Eling:2016xlx}), which corresponds to the case of an isolated horizon.}
	\begin{align}\label{app:symp-form-final}
		\begin{split}
			\O[g;\delta_1 g, \delta_2 g] = \frac{1}{8\pi G_N} \delta \a \wedge \delta A . 
		\end{split}
	\end{align}
	We have dropped the tilde on $\O$ as the above expression is invertible, and this is precisely Eq.~\eqref{eq:symp-form-final}.

\end{document}